\documentclass[%
  aps,prl,twocolumn,10pt,
  reprint
]{revtex4-2}

\usepackage{graphicx}
\usepackage{mwe}   
\usepackage{amsmath}
\usepackage{xcolor}
\usepackage{amsfonts}
\usepackage{booktabs} 
\usepackage{amssymb}
\usepackage{subcaption} 
\usepackage{caption}
\usepackage{makecell}
\usepackage{comment}
\usepackage{placeins}
\usepackage{wrapfig}

\usepackage[title]{appendix}
\usepackage{booktabs} 

\begin{document}

\title{Dense Packing of Tetrahedra in Cylinders}

\author{Gabrielle N. Jones}
\thanks{These authors contributed equally to this work.}
\affiliation{Department of Chemical Engineering, University of Michigan, Ann Arbor, MI 48109, USA}
\author{Adil Mughal}
\thanks{These authors contributed equally to this work.}
\affiliation{Department of Mathematics, Aberystwyth University, Aberystwyth, Wales SY23, UK}
\author{Philipp W.A. Schönhöfer}
\affiliation{Department of Chemical Engineering, University of Michigan, Ann Arbor, MI 48109, USA}
\author{Sharon C. Glotzer}
\affiliation{Department of Chemical Engineering, University of Michigan, Ann Arbor, MI 48109, USA}
\affiliation{Biointerfaces Institute, University of Michigan, Ann Arbor, MI 48109, USA}
\email{aqm@aber.ac.uk}
\email{gabs@umich.edu}

\begin{abstract}
We present the first systematic investigation of regular tetrahedra densely packed within a circular cylinder. Combining numerical simulations with exact analytical theory, we classify the densest configurations into two continuous, mathematically tractable families: chiral \textit{Tetra Helices} and achiral \textit{Dimer Chains}. Crucially, we reveal that the celebrated Boerdijk--Coxeter Helix is not an isolated anomaly, but a natural consequence of this cylindrical confinement. We further identify a striking symmetry-breaking bifurcation at a critical diameter, where the chiral helix coexists with an achiral dimer chain at a strictly identical packing fraction. 
\end{abstract}

\maketitle

\begin{table*}[!ht]

\centering
\caption{Gallery and classification of the six densest tetrahedron-in-cylinder structures observed in the range $1.0 \leq D/a \leq 1.21045$. The structures are identified by packing number, name, range of $D/a$ over which they occur and packing fraction.}
\label{tab:gallery_of_images}

\setlength{\tabcolsep}{5pt}
\renewcommand{\arraystretch}{1.25}

\includegraphics[width=0.8\textwidth]{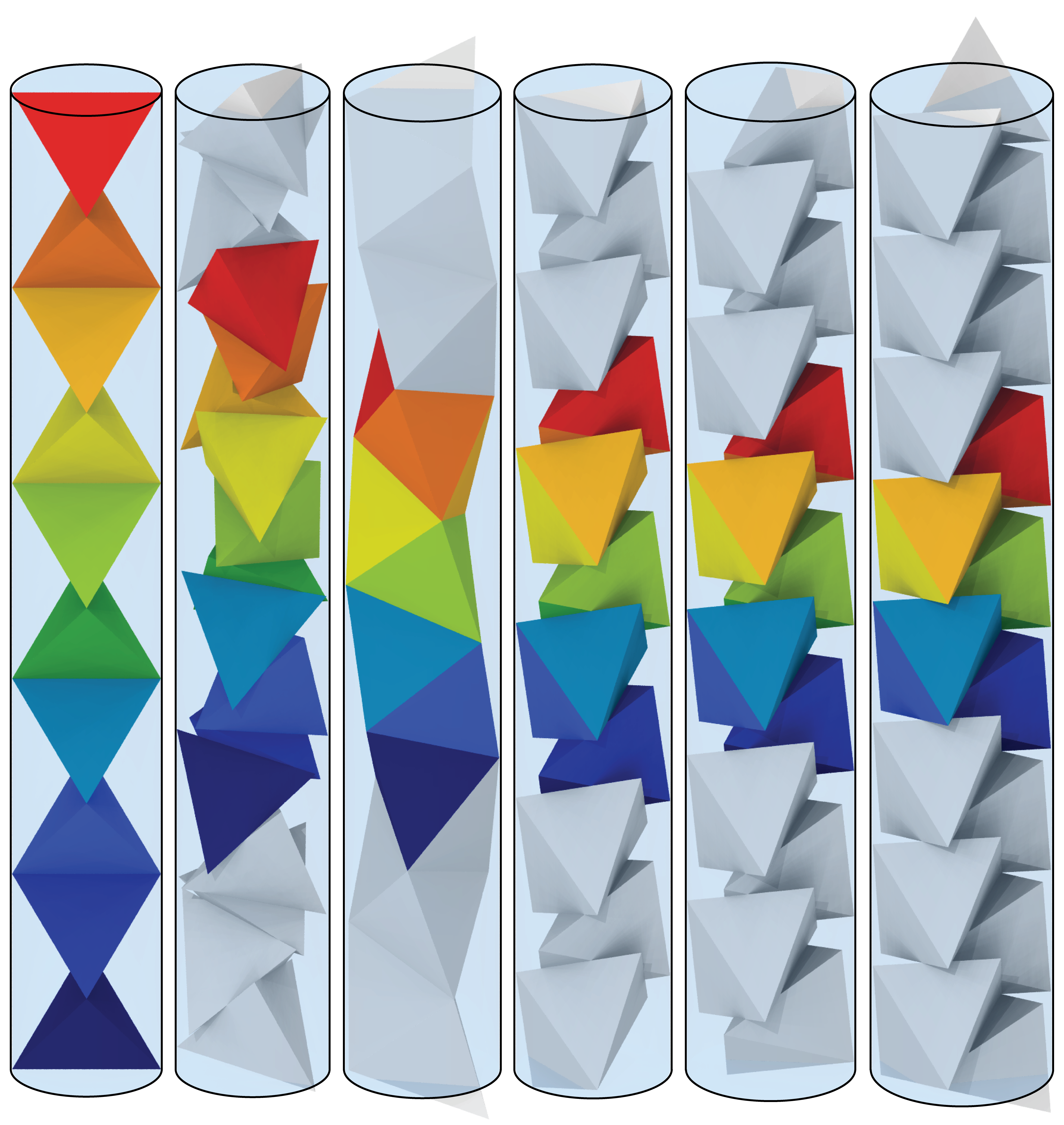} \\

\begin{tabular}{cccccc}
\toprule

\multicolumn{6}{c}{\textbf{Packing Number}}
\\[0.2em]
1 & 2 & 3 & 4 & 5 & 6
\\[0.4em]

\midrule

\multicolumn{6}{c}{\textbf{Name}}
\\[0.2em]
\makecell{Edge-Sharing\\ Tetra Chain}& 
\makecell{Face-Sharing\\ Tetra Helix}& 
\makecell{Boerdijk--Coxeter\\ Tetra Helix}& 
\makecell{Maximal-Contact\\ Dimer Chain}& 
\makecell{Open\\ Dimer Chain}& 
\makecell{Compact\\ Dimer Chain}
\\[0.4em]

\midrule

\multicolumn{6}{c}{\textbf{Range}}
\\[0.2em]
\makecell{$\{1\}$}& 
\makecell{$\left[1,\frac{3\sqrt{3}}{5}\right]$}& 
\makecell{$\{\frac{3\sqrt{3}}{5}\}$}& 
\makecell{$\{\frac{3\sqrt{3}}{5}\}$}& 
\makecell{$\left[\frac{3\sqrt{3}}{5},\,1.21045\right]$}& 
\makecell{$\{1.21045\}$}
\\

\midrule

\multicolumn{6}{c}{\textbf{Packing Fraction}}
\\[0.2em]
\makecell{$\frac{2}{3\pi}$}& 
\makecell{See Eq. \eqref{eq:fs_density}}& 
\makecell{$\frac{50\sqrt{5}}{81\pi}$}& 
\makecell{$\frac{50\sqrt{5}}{81\pi}$}& 
\makecell{See Eq. \eqref{eq:dc_density}}& 
\makecell{$0.465846$}
\\

\bottomrule
\end{tabular}

\end{table*}

Packing problems play a central role in mathematics and physics, with implications spanning fundamental geometry and materials science \cite{weaire2008pursuit, chan2025packing}. The quest to determine the densest possible arrangements of congruent objects has led to breakthroughs ranging from Kepler’s conjecture on sphere packings \cite{hales2011historical} to modern challenges in self-assembly and nanotechnology \cite{manoharan_colloidal_2015}.

While much of this work has focused on dense bulk packings, namely the densest arrangements of objects in unbounded space, many physical \cite{ouguz2012packing, vanapalli2008fluidic, kim2022designing} and biological systems \cite{berndsen2014nonequilibrium, tallinen2016growth} instead involve the dense organization of particles, filaments, or extended material within confined spaces, constrained by walls, pores, channels, membranes, or other confining geometries. Packing under confinement has therefore emerged as a distinct problem \cite{pickett2000spontaneous, marenduzzo2010biopolymer, chen2015confined, wang2018interplay}, but remains far less developed than its bulk counterpart.

Confinement can give rise to especially rich packing behavior, as local geometric constraints compete with global organization to produce unexpected structures \cite{wan_shapes_2018, pickett2000spontaneous, ouguz2012packing, mughal2012dense,teich2016clusters}. In this Letter we present the first systematic results on the dense packing of congruent regular tetrahedra inside a circular cylinder, using the ratio $D/a$ of the cylinder diameter $D$ to the tetrahedral edge length $a$ as the controlling parameter. 

Remarkably, we find that these packings are amenable to analysis: they organize into families of chiral/achiral, screw periodic, structures distinguished by the number of tetrahedral vertices in contact with the cylindrical boundary. We study the range $1 \leq D/a \leq 1.21045$ and identify two families of dense tetrahedral packings, which we call \emph{Tetra Helices} and \emph{Dimer Chains}. In this setting, the celebrated Boerdijk\cite{boerdijk_remarks_1952}--Coxeter\cite{coxeter1985simplicial} helix emerges not as an isolated structure, but as one member of a broader tetrahelix sequence selected by cylindrical confinement.

The present results bring together two previously distinct lines of research: the dense packing of regular tetrahedra\cite{conway_packing_2006, lagarias_mysteries_2012, wang_structural_2022, wang_controlled_2023} and the dense packing of particles in cylinders. In the bulk, regular tetrahedra are known to form structures built from tetrahelical and dimer motifs. Tetrahelical arrangements appear within dodecagonal quasicrystalline phases of hard tetrahedra \cite{haji-akbari_disordered_2009} and dense hypersphere packings \cite{schonhofer_rationalizing_2023}, while the densest known bulk packings are crystalline arrangements based on face-sharing tetrahedral dimers \cite{chen_dense_2010}. The recurrence of these motifs in both bulk and confined settings suggests that tetrahelical chains and face-sharing dimers are fundamental organizing principles for dense tetrahedral packing.

Cylinder-confined packing, by contrast, has been developed most extensively for hard spheres. In that setting, the controlling parameter is the diameter ratio $D/d$ between the cylinder diameter $D$ and the sphere diameter $d$ \cite{pickett2000spontaneous, mughal_phyllotactic_2011, mughal2012dense, fu_assembly_2017}. Varying $D/d$ generates a sequence of quasi-one-dimensional columnar crystals, including chiral helical packings with no direct analogue in bulk three-dimensional sphere packings. Our work combines these two perspectives by asking how the orientational constraints and local contact motifs of regular tetrahedra are organized by the quasi-one-dimensional geometry imposed by a cylinder.

For arbitrary values of the cylinder-to-particle size ratio $D/a$, optimal packings of hard tetrahedra cannot in general be inferred \emph{a priori} and were therefore sought numerically. Following the geometric framework used in previous studies of columnar sphere packings \cite{mughal_phyllotactic_2011, mughal2012dense, fu_assembly_2017} (and spheroids \cite{jin2020shape}), we restricted our search to periodic packings with screw symmetry. Each candidate packing is generated from a primitive cell of axial length $L$ containing $N$ tetrahedra; the full structure is obtained by repeating this cell under a combined axial translation by $L$ and azimuthal rotation by a twist angle $\alpha$. Dense candidate packings were explored by varying $(L,N,\alpha)$ together with the positions and orientations of the tetrahedra within the primitive cell, for $1 \leq N \leq 20$.

The numerical search was carried out using HOOMD-blue \cite{anderson_hoomd-blue_2020} with a pressure-driven compression protocol designed to identify packings with high packing fractions under cylindrical confinement. Starting from dilute initial conditions, candidate structures were compressed and relaxed while enforcing the confinement geometry and non-overlap of the final configurations, after which the resulting cell length and packing fraction were recorded. A detailed summary of the simulation method is provided in the supplementary material \cite{supp}.  

\begin{figure}
    \centering
    \includegraphics[width=0.9\linewidth]{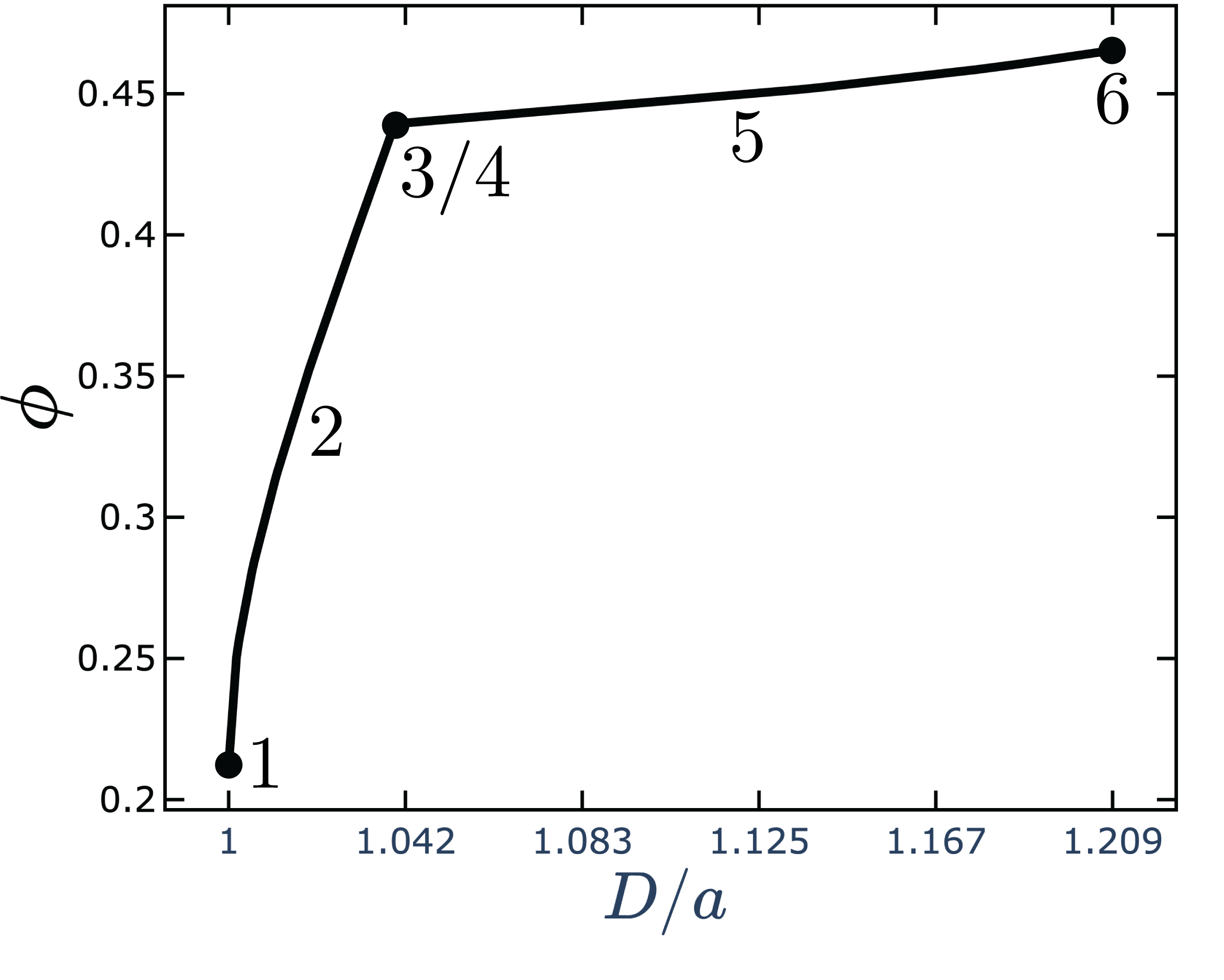}
    \caption{Packing fraction $\phi$ of analytical constructions for dense tetrahedral packings as a function $D/a$. Labels denote the structures defined in Table \ref{tab:gallery_of_images}. The sequence bifurcates into two distinct structural regimes: the chiral Tetrahelices (structures 1--3) and achiral Dimer-Chains (structures 4--6).}
    \label{fig:densitiesFullRangeWithAnalytics}
\end{figure}

Our results are summarized in Table~\ref{tab:gallery_of_images}, which lists the six densest structures identified over the range $1.0 \leq D/a \leq 1.21045$ along with their representative images of these packings. The packing fractions as functions of $D/a$ are plotted in Fig.~\ref{fig:densitiesFullRangeWithAnalytics}. Lower packing fraction ordered structures found in simulation are presented in the supplementary material, without analytical constructions \cite{supp}. Animations showing smooth, continuous  transitions between structures 1-3 and 3-6 are also presented in the supplementary material. We now discuss this sequence of densest packings in turn.

\noindent{\bf 1. Edge-Sharing Tetra Chain:}  Structure 1 (see Table~\ref{tab:gallery_of_images}) occurs at $D/a = 1$ and $N=1$. Each tetrahedron is oriented with a pair of mutually perpendicular edges normal to the cylinder axis. The centroid of the tetrahedron lies on the cylindrical axis and the four vertices of the tetrahedron touch the cylindrical boundary.

The packing is achiral and has a unit cell of axial length $L=a/\sqrt{2}$ containing a single tetrahedron, and packing fraction $\phi=\frac{2}{3\pi}\approx 0.212207$.

The Edge-Sharing Tetra Chain is naturally identified as the $\alpha=\pm\pi/2$ limiting case of structure 2 (see below). However, the packing may equally be viewed as a rotor in which each tetrahedron has a continuous freedom to rotate about the $z$-axis.

\noindent{\bf 2. Face-Sharing Tetrahelix:} For $D/a>1$ and $N=2$, The Edge-Sharing Tetra Chain continuously deforms into a chiral Face-Sharing Tetrahelix (structure 2 in Table~\ref{tab:gallery_of_images}). 

Neighboring tetrahedra make \emph{partial} face-to-face contact, with the overlap region increasing continuously as the cylinder diameter increases. This progressively compacts the packing, leading to a corresponding increase in packing fraction, as shown in Fig.~\ref{fig:densitiesFullRangeWithAnalytics}. 

The unit cell consists of a single tetrahedron translated along the cylinder by a distance
\begin{equation}
\label{eq:fs_length}
\frac{L(\theta)}{a}=\frac{1-\cos\theta\sin\theta}{\sqrt{2}(\cos\theta+\sin\theta)},
\end{equation}
and rotated around the cylindrical axis by a twist angle
\begin{equation}
\label{eq:fs_twist}
\alpha = \chi \left( \arctan\left( \csc\theta - \sec\theta\right)-\pi \right), \;\;\ \chi=\pm1,
\end{equation}
where the choice of $\chi$ corresponds to the two chiral enantiomers, and $\theta \in (0,\arctan[\frac{1}{2}])$ is a dimensionless parameter. 

All four vertices of the tetrahedron remain in contact with the cylindrical boundary. For each admissible value of $D/a$, there is a unique solution, up to a screw displacement, that satisfies this four-vertex contact condition. Defining
\begin{equation*}
\begin{aligned}
\mu &:= \frac{(2+\sin 2\theta)}{4\sqrt{2}}, &
\nu &:= \frac{(2-\sin 2\theta)}{4\sqrt{2}}, \\
\sigma &:= \frac{(\cos\theta+\sin\theta)}{2\sqrt{2}}, &
\tau &:= \frac{(\cos\theta-\sin\theta)}{2\sqrt{2}},
\end{aligned}
\end{equation*}
then (up to a screw displacement) the tetrahedron vertices are given by
\begin{equation}
\begin{aligned}
\mathbf{v}_1 &= a(-\mu,\phantom{-}\chi\tau,\phantom{-}\sigma),\\
\mathbf{v}_2 &= a(\phantom{-}\nu,-\chi\sigma,\phantom{-}\tau),\\
\mathbf{v}_3 &= a(\phantom{-}\nu,\phantom{-}\chi\sigma,-\tau),\\
\mathbf{v}_4 &= a(-\mu,-\chi\tau,-\sigma).
\label{eq:tetraHelixVertices}
\end{aligned}
\end{equation}

The radius \(r = |\mathbf{v}_i|\) of the confining cylinder can be computed from Eq.~\eqref{eq:tetraHelixVertices}, yielding
\begin{equation}
\label{eq:fs_radius_diameter}
\frac{D(\theta)}{a} = 2\frac{r(\theta)}{a} = \sqrt{1 + \frac{1}{8}\sin^2 2\theta}.
\end{equation}

Hence, the packing fraction of the Face-Sharing Tetrahelix packing is given by 
\begin{equation}
\phi(\theta)=\frac{V_T}{\pi r(\theta)^2 L(\theta)} 
\label{eq:fs_density}
\end{equation}
where $V_T=a^3/(6\sqrt{2})$ is the volume of a tetrahedron of side length $a$. 

In the limit $\theta \to 0$, the Face-Sharing Tetrahelix reduces to the Edge-Sharing Tetra Chain, while in the upper limit $\theta \to \arctan\left(\frac{1}{2}\right)$, it converges to the Boerdijk--Coxeter Tetrahelix (see below).

\noindent{\bf 3. Boerdijk--Coxeter Tetrahelix:} In the limit of \emph{complete} face-to-face contact, the Face-Sharing tetrahelix family culminates in the celebrated Boerdijk--Coxeter Helix \cite{boerdijk_remarks_1952, coxeter1985simplicial}. Viewed structurally, this limiting case forms a chiral sequence of regular tetrahedra where each successive tetrahedron is in full contact with the next tetrahedron in the chain (structure 3 in Table~\ref{tab:gallery_of_images}).

Imposing the known cylinder radius, i.e. setting $r=\frac{3\sqrt{3}}{10}a$ in Eq. \eqref{eq:fs_radius_diameter} and solving for $\theta$, gives $\theta= \arctan\left(\frac{1}{2}\right)$. Substituting this value of $\theta$ into Eqs. \eqref{eq:fs_length} and \eqref{eq:fs_twist},  yields
\[
\frac{L}{a} = \frac{1}{\sqrt{10}},
\qquad
\alpha = \pm\arccos\left(-\frac{2}{3}\right)
= \pm 2\arctan\left(\sqrt{5}\right),
\]
and also $D/a = (3\sqrt{3})/5$, which are precisely the standard geometric parameters of the Boerdijk--Coxeter tetrahelix. The structure consists of a single tetrahedron per unit cell and the resulting packing fraction is
\[
\phi=\frac{50\sqrt{5}}{81\pi},
\]
as plotted in Fig.~\ref{fig:densitiesFullRangeWithAnalytics}.

\noindent \textbf{4. Maximal-Contact Dimer Chain:}
Remarkably, at $D/a=3\sqrt{3}/5$ and $N=4$ we find a second packing with the same packing fraction as the Boerdijk--Coxeter Tetrahelix. We call this structure the \emph{Maximal-Contact Dimer Chain}. The coexistence of these two distinct packings is marked by the point labeled 3/4 in Fig.~\ref{fig:densitiesFullRangeWithAnalytics}. The two structures therefore have the same packing fraction at the same confinement ratio, but are geometrically distinct packings rather than configurations connected by a continuous deformation at fixed $D/a$.

\begin{figure}
    \includegraphics[width=\linewidth]{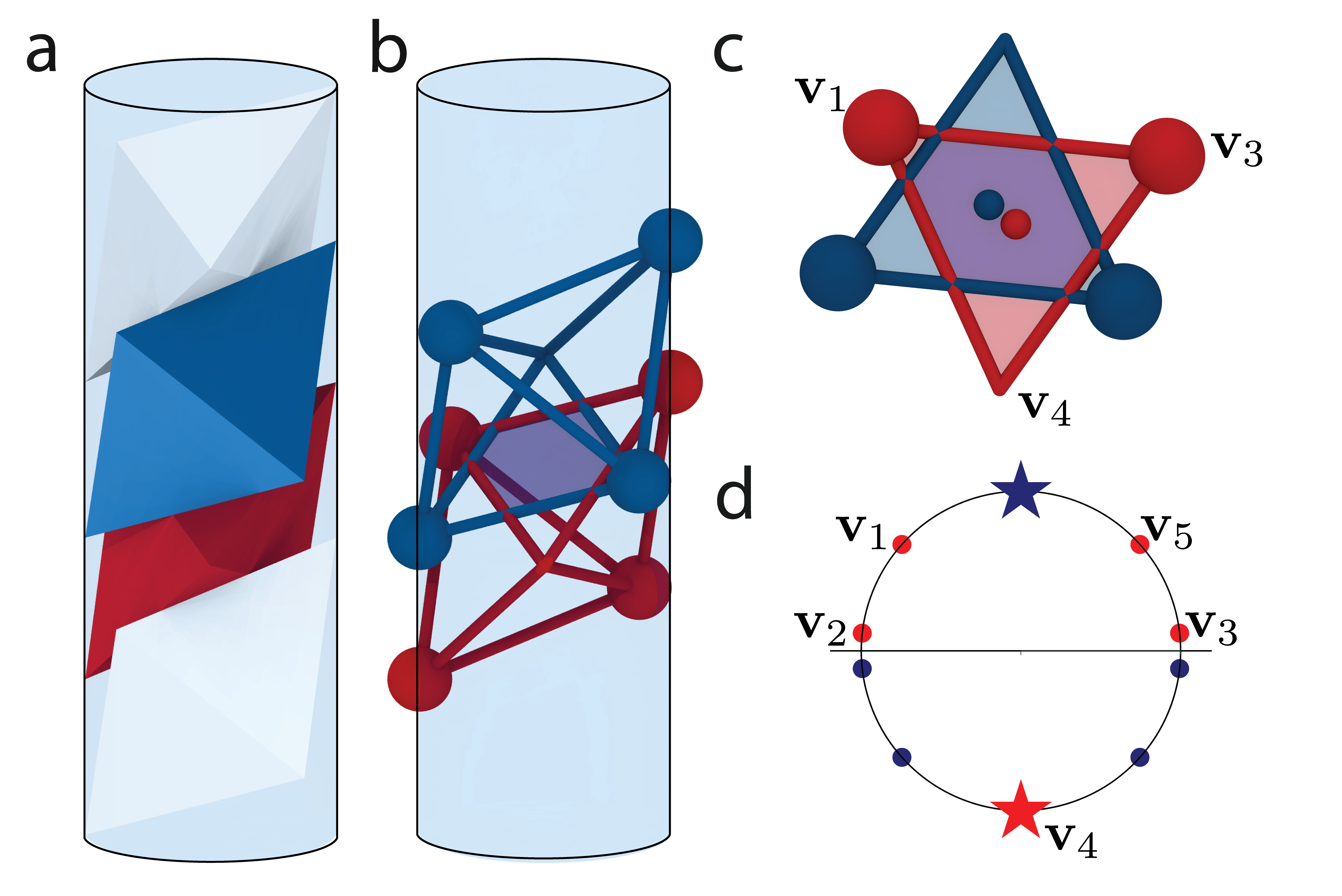}
   \caption{Geometric construction of the Maximal-Contact Dimer Chain. 
(a) Two adjacent dimers, each a triangular bipyramid. 
(b) Wire-frame view with boundary-contact vertices $\mathbf{v}_1, \mathbf{v}_2, \mathbf{v}_3, \mathbf{v}_5$ marked by spheres. 
(c) Contacting faces between neighboring dimers, viewed along the face normal, showing offset face centroids. 
(d) Projection along the cylinder axis showing the mirror plane of the glide reflection; starred vertices lose wall contact as $D/a$ is increased.}
    \label{fig:maximalContact}
\end{figure}

The packing consists of a stacked chain of dimers, each formed from two tetrahedra sharing a complete triangular face, equivalently a triangular bipyramid; see structure 4 in Table~\ref{tab:gallery_of_images} and Fig.~\ref{fig:maximalContact}a. In this limiting configuration, all four vertices of each tetrahedron lie on the cylindrical boundary, motivating the designation ``Maximal-Contact''. Their coordinates are therefore given by the all-contact parametrization in Eq.~\eqref{eq:tetraHelixVertices}, with
\[
\theta=\arctan\left(\frac{1}{2}\right),
\]
corresponding to $D/a=3\sqrt{3}/5$. The two tetrahedra comprising each dimer are consequently related by the standard Boerdijk--Coxeter screw displacement.

A wire-frame representation of two neighboring dimers is shown in Fig.~\ref{fig:maximalContact}b. In the Maximal-Contact Dimer Chain, all five vertices of each triangular bipyramid remain in contact with the cylindrical boundary. This property is special to this limiting structure. As the packing opens into the Dimer Chain family described below, one vertex of each dimer (labeled $\mathbf{v}_4$ below) loses contact with the cylinder, while the other four remain on the cylindrical boundary. In Fig.~\ref{fig:maximalContact}b these persistent boundary-contact vertices are marked by spheres, while the vertex (i.e. $\mathbf{v}_4$) that subsequently loses contact is left undecorated; the same vertex is starred in the axial projection in Fig.~\ref{fig:maximalContact}d.

Fig.~\ref{fig:maximalContact}c shows the shared faces between two neighboring dimers, viewed along the face normal, with boundary-contact vertices highlighted as in Fig.~\ref{fig:maximalContact}b. The centroids of these contacting faces are offset rather than coincident. Successive dimers are related by a glide reflection, as shown in Fig.~\ref{fig:maximalContact}d, i.e. a reflection across the indicated mirror plane followed by a translation along the central axis. The required axial translation is $L/a=\sqrt{10}/5$, which gives the spacing at which neighboring dimers meet in face-to-face contact.

The equality in packing fraction with the Boerdijk--Coxeter Tetrahelix follows immediately from the axial period. The Boerdijk--Coxeter helix packs one tetrahedron into a cylinder segment of length $L/a=1/\sqrt{10}$, whereas the Maximal-Contact Dimer Chain packs two tetrahedra, i.e. one dimer, into a unit cell of twice this length.

\noindent \textbf{5. Open Dimer Chain:}
$D/a = \frac{3\sqrt{3}}{5}$ marks a symmetry-breaking bifurcation in the packing sequence: the chiral screw-symmetric Tetrahelix family gives way to the achiral glide-symmetric Dimer Chain packings.

For $D/a > \frac{3\sqrt{3}}{5}$ and $N=4$, the maximal-contact dimer chain continuously deforms into the achiral Open Dimer Chain. The dimers begin to ``slide'' apart with increasing $D/a$, forming a more pronounced zigzag arrangement (structure 5 in Table~\ref{tab:gallery_of_images}). This allows the packing to compact further, yielding the continuous increase in packing fraction observed in Fig.~\ref{fig:densitiesFullRangeWithAnalytics}. 

This shift is accommodated by one vertex of each dimer losing contact with the cylindrical boundary; in the Open Dimer Chain, this vertex lies in the interior of the cylinder, as indicated by the starred vertex in Fig.~\ref{fig:maximalContact}d. At the same time, the dimers rotate into a flatter orientation, becoming more closely aligned with the plane transverse to the cylinder axis.

We impose two geometric constraints: the remaining four exterior vertices of each dimer must remain on the cylindrical surface, and successive dimers must maintain face-to-face contact via a glide reflection. Solving these constraints reveals that the geometry of the Open Dimer Chain is governed by a single dimensionless parameter $\lambda/a$.

The diameter of the confining cylinder is given by
\begin{equation}
    \frac{D(\lambda)}{a}=2\frac{r(\lambda)}{a} = 2\sqrt{\frac{(\lambda/a)^2+72}{363}}.
    \label{eq:f1Rovera}
\end{equation}
Introducing the shorthand

\begin{equation}
\omega_1=\sqrt{7+\left(\frac{\lambda}{a}\right)},
\quad
\omega_2=\sqrt{93-4\left(\frac{\lambda}{a}\right)},
\quad
\omega_3=\left(\frac{\lambda}{a}\right)-4,
\label{eq:f1OmegaDefs}
\end{equation}
where $\lambda/a$ is restricted to the range $51/10 \le \lambda/a \lesssim 7.808$ (see below), we find that the spacing between successive dimers is given by
\begin{equation}
    \frac{L(\lambda)}{a}
    =
    \frac{2\sqrt{2}(11-\omega_3)}
    {6\sqrt{2}\,\omega_1+\sqrt{3}\,\omega_2}.
    \label{eq:f1hzovera}
\end{equation}
The packing fraction of the construction is found to be
\begin{equation}
\label{eq:dc_density}
\phi(\lambda)=\frac{2V_T}{\pi r(\lambda)^2 L(\lambda)}. 
\end{equation}
The position of the dimer vertices is 
\begin{equation}
\begin{aligned}
\mathbf{v}_1 &= a(\phantom{-}\alpha,\phantom{-}\beta,\phantom{-}\gamma),\\
\mathbf{v}_2 &= a(\phantom{-}\delta,\phantom{-}\epsilon,-\zeta),\\
\mathbf{v}_3 &= a(-\delta,\phantom{-}\epsilon,\phantom{-}\zeta),\\
\mathbf{v}_4 &= a(\phantom{-}0,\phantom{-}\eta,\phantom{-}0),\\
\mathbf{v}_5 &= a(-\alpha,\phantom{-}\beta,-\gamma),
\end{aligned}
\label{eq:f1VZDefinitions}
\end{equation}
where, 
\begin{align}
\alpha &= -\frac{2\sqrt{6}}{33}\omega_1,
&\hspace{-0.8em}
\beta &= \frac{\sqrt{3}}{33}\omega_3,
&\hspace{-0.8em}
\gamma &= -\frac{\sqrt{6}}{33}\omega_2,
\notag\\
\delta &= -\frac{1}{22}\omega_2,
&\hspace{-0.8em}
\epsilon &= \sqrt{3}\!\left(\frac{1}{6}+\frac{\omega_3}{33}\right),
&\hspace{-0.8em}
\zeta &= -\frac{1}{11}\omega_1,
\notag\\
\eta &= \frac{\sqrt{3}}{33}(\omega_3-11).
\label{eq:f1ABCOmega}
\end{align}

Letting $\Delta z/a$ represent the internal axial offset between the two tetrahedra within a single dimer,  we find that $L/a < 2(\Delta z/a)$ for the Open Dimer Chain. This is in contrast to the Maximal Contact case, where $L/a = 2(\Delta z/a)$.

The Open Dimer Chain configuration is physically valid only within a specific domain of $D/a$. To prevent physical intersection, the parameter $\lambda/a$ is restricted to the range given above, meaning the diameter ratio $D/a$ is bounded within the range
\begin{equation} \label{eq:od_domain}
    \frac{3\sqrt{3}}{5} \le \frac{D}{a} \lesssim 1.21045.
\end{equation}
The lower bound corresponds to the Maximal Contact limit. Above the upper threshold, explained in the supplementary material \cite{supp}, this construction forces unphysical overlaps between successive dimers, rendering the packing family physically invalid.

\noindent \textbf{6. Compact Dimer Chain:}
At the upper boundary $D/a \approx 1.21045$, next-nearest dimers in the Open Dimer Chain come into direct contact, as shown by structure 6 in Table~\ref{tab:gallery_of_images}. This limiting structure has packing fraction $\phi=0.465846$ and is termed the \textit{Compact Dimer Chain}: the gaps between next-nearest neighbors have closed, so any further increase in cylinder diameter would produce unphysical overlaps.

\noindent{\bf Conclusions:}
We have presented the first systematic numerical and analytical study of dense packings of regular tetrahedra in circular cylinders, a problem whose richness has recently been anticipated in broader discussions of confined packing \cite{chan2025packing}. Even over the restricted range studied here, $1\leq D/a\leq1.21045$, cylindrical confinement generates a nontrivial sequence of structures, organized into Tetra Helix and Dimer Chain families. The Boerdijk--Coxeter tetrahelix appears not as an isolated packing, but as one member of a broader sequence. A notable outcome is the identification of a previously unknown achiral Maximal-Contact Dimer Chain, which remarkably has exactly the same packing fraction as the chiral Boerdijk--Coxeter Tetra-Helix (and the same value of $D/a$).

These results suggest that strong confinement can turn otherwise complex three-dimensional packing problems into controlled quasi-one-dimensional geometries, in which dense polyhedral packings may be described analytically. How far these analytic constructions can be continued as $D/a$ is increased remains open. For instance, simulations beyond the $D/a>1.21045$ limit suggest, although not analytically proven, that other complex structures, such as pentagonal stacks, can achieve ideal packing in cylindrical confinement. These structures are reminiscent of the pentagonal columns observed in the dodecagonal quasicrystal formed by hard tetrahedra in bulk \cite{haji-akbari_disordered_2009}, however feature a 10-fold instead of a 12-fold ring arrangement (see Supplementary Material \cite{supp}). Nevertheless, the present work indicates that similar methods may be applicable to the dense packing of other Platonic solids in cylinders.

The broader significance lies in the controlled emergence of chiral and achiral structures from achiral building blocks. Natural chiral biomaterials, including iridescent beetle carapaces \cite{mendoza-galvan_graded_2018}, phyllotactic spirals \cite{douady_phyllotaxis_1992}, hierarchical bone structures \cite{zhou_chiral_2022a}, and twisted collagen bundles \cite{emam_collagen_2025}, show how geometric chirality can be coupled to optical, mechanical, and structural function. They have therefore inspired the design of synthetic chiral materials \cite{kumar_photonically_2023, hall_morphology_2016, grason_chiral_2020, liu2023chiral}. In this context, the confined tetrahedron packings described here provide a simple geometric setting in which confinement alone selects between chiral and achiral dense structures, suggesting possible design principles for self-assembled photonic, plasmonic, and mechanically tunable materials.

\noindent{\bf Acknowledgments: } The authors would like to thank the Isaac Newton Institute for Mathematical Sciences, Cambridge, for support and hospitality during the program \textit{5th International Conference on Packing Problems: Packing and patterns in granular mechanics}, where work on this paper was undertaken. This work was partially supported by EPSRC grant EP/R014604/1. GNJ acknowledges support from the U.S. Department of Energy, Office of Science, Office of Advanced Scientific Computing Research, Department of Energy Computational Science Graduate Fellowship under Award Number DE-SC0022158. This work used Anvil at Purdue RCAC through allocation DMR 140129 from the Advanced Cyberinfrastructure Coordination Ecosystem: Services \& Support (ACCESS) program, which is supported by National Science Foundation grants \#2138259, \#2138286, \#2138307, \#2137603, and \#2138296.  Computational resources and services were also provided by Advanced Research Computing (ARC), a division of Information and Technology Services (ITS) at the University of Michigan, Ann Arbor. PWAS and SCG acknowledge support by a Vannevar Bush Faculty Fellowship sponsored by the Department of the Navy, Office of Naval Research under ONR award number N00014-22-1-2821.

\bibliography{CylinderBib}

\clearpage
\appendix

\onecolumngrid

\begin{center}
    {\Large\bfseries Supplementary Information}
\end{center}

\section{Simulation Protocol}

\begin{figure}[h]
    \centering
    \includegraphics[width=\linewidth]{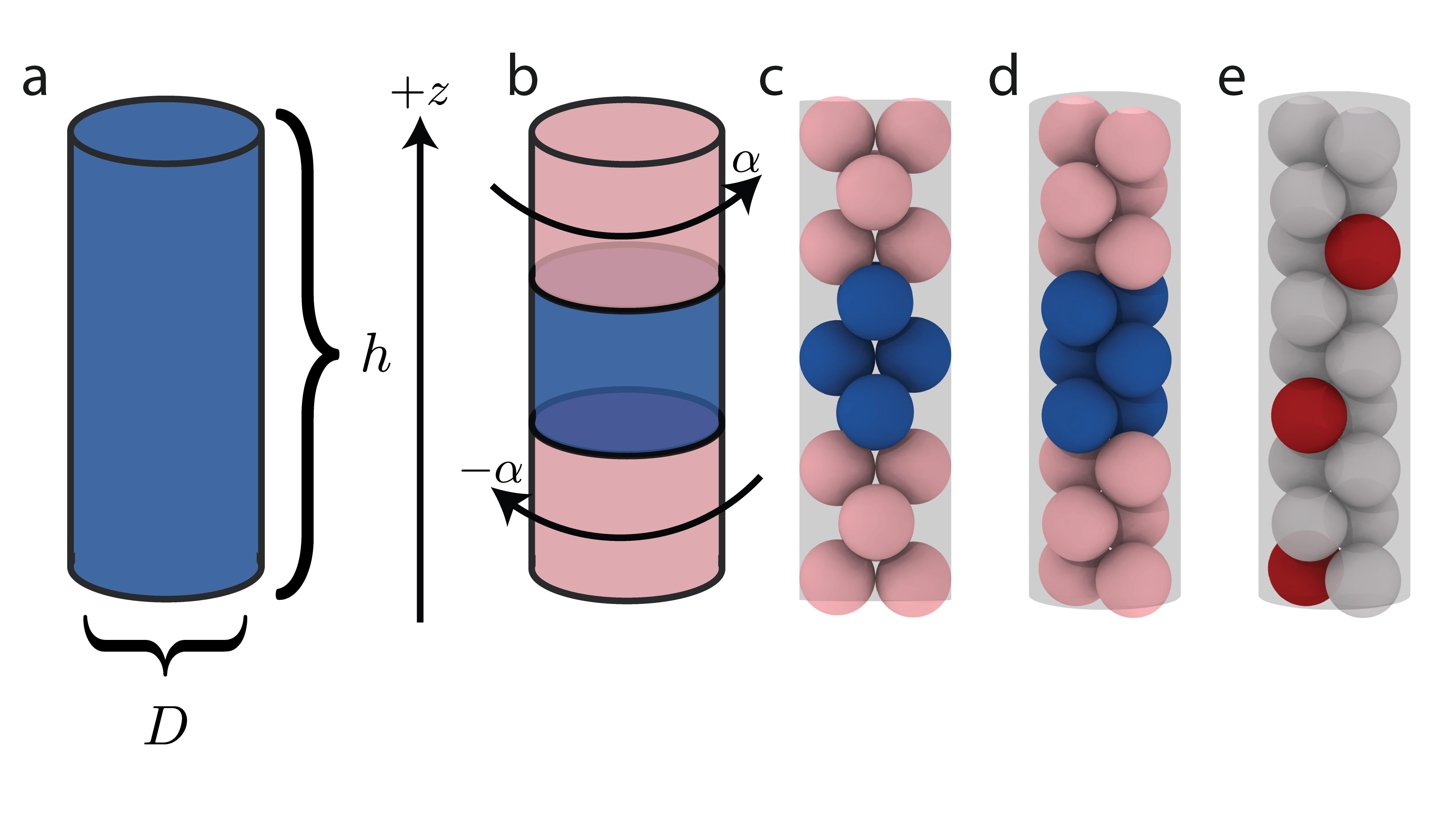}
    \caption{Simulation parameters: (a) The confining cylinder is defined by a diameter $D$ and periodic height $h$. (b) Each periodic image in the $\pm z$ direction is rotated by a twist angle $\pm \alpha$ about the z-axis. (c, d) A representative image of 6 spheres in a cylinder with dimensionless diameter $D/d = 2$ colored by the primary cell in blue and periodic images in pink. Here, $\alpha = \pi/2$. (e) Highlighting the same particle in each subsequent image in red demonstrates this twist angle $\alpha=\pi/2$.}
    \label{fig:cylinder_parameters}
\end{figure}

We ran HPMC simulations within cylindrical confinement using a modified version of the HOOMD-blue simulation toolkit (version 5.2.0), run on Anvil\footnote{Gerry McCartney, Thomas Hacker, and Baijian Yang. Empowering Faculty: A Campus Cyberinfrastructure Strategy for Research Communities. \textit{Educause Review}, 2014}, a high performance computing cluster at Purdue University. All particles remain within the interior of a cylindrical boundary, which has rotational symmetry along the $z$-axis and diameter $D$. Platonic tetrahedra are described using their edge length $a$. To simulate infinite systems, which may have a unit cell that repeats with some twist about the cylindrical axis to form a chiral structure, we implement a twisted periodic boundary condition. This condition rotates the position and orientation of a particle about the cylindrical $\pm$z-axis in each periodic image by a twist angle $\pm\alpha$. All of these cylinder parameters are visually represented in figure \ref{fig:cylinder_parameters}.\\

\begin{figure}[h]

    \includegraphics[width=0.39\textwidth]{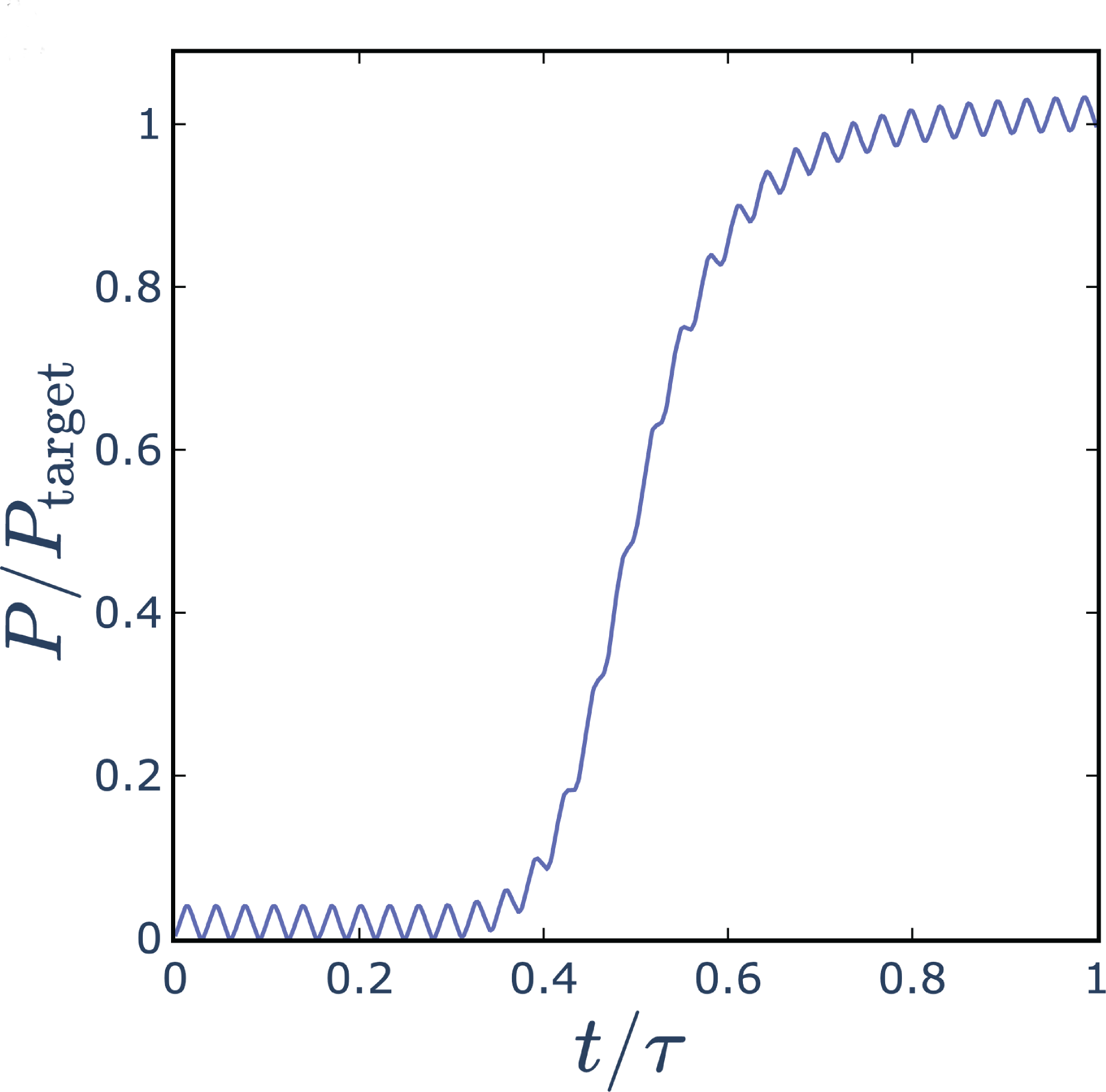}
    \caption{Pressure variant: The pressure $P$ as a function of MC sweep $t$. Values are normalized by the target pressure $P_\textrm{target}$ and the total number of MC sweeps $\tau$.}
    \label{fig:cylinder_pressure_variant}
\end{figure}

Although the ideal unit cell of the system will have the minimal number of particles possible, this is not known \textit{a priori}. Given this, we need to examine a range in the number of particles. We initialize $1\le N \le 20$ evenly spaced particles in the cylinder with height $h = 20nR_\textrm{CS}$ where $R_\textrm{CS}$ is the circumsphere radius of the particle. The cylinder height $h$ is compressed to a target pressure\footnote{Enrique De Miguel and George Jackson. Detailed examination of the calculation of the pressure in simulations of systems with discontinuous interactions from the mechanical and thermodynamic perspectives. \textit{Molecular Physics}, 104(22-24):3717–3734, November 2006} ($P_{\textrm{target}}$) using a ramp function over $\tau = 10^6$ MC sweeps with form $P(t, \tau) = \left(1+e^{-10t/\tau}\right)^{-1} + 0.02*\sin(200t/\tau - \pi/2) + 0.02$, seen in figure \ref{fig:cylinder_pressure_variant}. We found that a sigmoidal increase in pressure is optimal when compared to step, linear, and logarithmic increases. The system is run at this high pressure for $2*10^{7}$ MC trial moves. After this stage, trial moves now allow temporary overlaps while further compressing the system to minimize cylinder length $h$ over $\tau = 5*10^{6}$ MC sweeps. For each value of $1 \le N \le 20$ we run 50 replicates and report the highest packing density per $n$. Simulations are run over a variety of cylinder diameters $1 \le D/a \lesssim 1.217$. For each combination $D$ and $a$ we report the the highest density, the number of particles in the unit cell $N_\textrm{max}$, and the corresponding height $h_\textrm{max}$. We denote the final packing density as $\phi = N_\textrm{max} v_{\textrm{particle}}/\left(\pi(D/2)^{2}h_\textrm{max}\right)$.

\clearpage
\section{Lower Density Structures}

\begin{figure}[h]
    \centering
    \includegraphics[width=\linewidth]{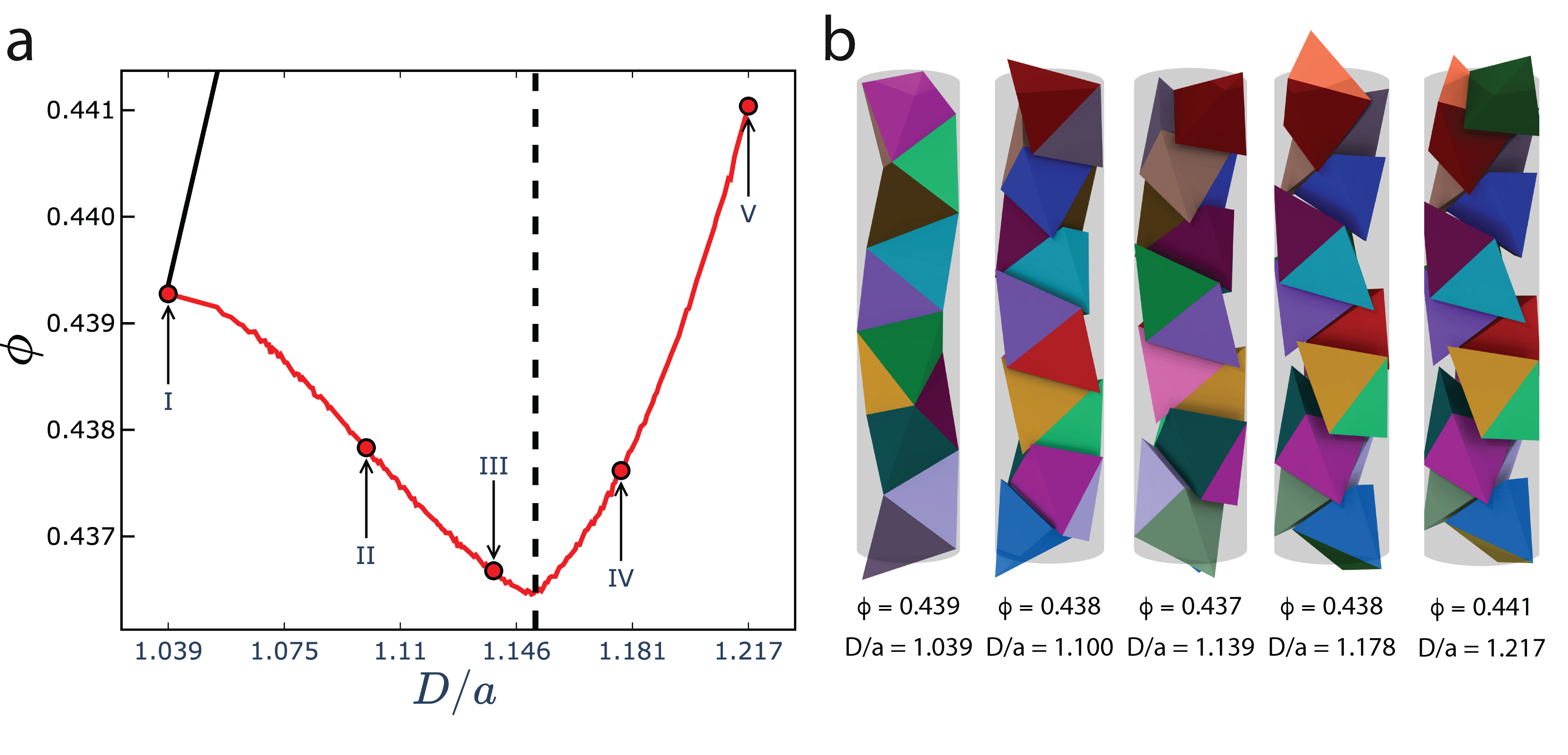}
    \caption{Suboptimal assemblies of regular tetrahedra: (a) The density of suboptimal assemblies of regular tetrahedra (red line) in cylinders larger than the diameter corresponding to the Boerdijk-Coxeter tetrahelix ($D/a\sim1.039$). The black curve corresponds to the Maximal Contact Dimer Chain. (b) (left to right) Five structures corresponding to points I-V in density plot (a). Note that point 1 is the Boerdijk-Coxeter tetrahelix. We observe a twisted chiral morphology in these structures with two particles in the unit cell.}
    \label{fig:lower_density}
\end{figure}

We find two lower density structures within the same diameter range of the maximal, open, and compact dimer chains when performing simulations with $N=2$. We show the density and morphology of these chiral structures in figure \ref{fig:lower_density}. Structures to the left of the dashed vertical line do not have dimer pairs, while those to the right of this line are formed with dimer pairs. Although structures I-III appear to have tetrahedral dimers, there is a small angular gap between pairs of tetrahedra. This is outside of the scope of this work, but is of interest to the self-assembly community at large.

\clearpage
\section{Highest Constructed Densities in terms of cylinder diameter $D$ and edge length $a$}

There is a threshold value of $D/a = 2R/a$ above which the construction of the Open Dimer Chain will lead to overlap between particles, generating invalid configurations. We find this upper bound, termed the Compact Dimer Chain, by finding the radius $R$ where any pair of edges between each dimer intersect. After substituting all vertex component values and finding the height between dimers $L$ we have an equation with form $f(R/a) = 0$. 

\begin{equation}
\begin{split}
    &f(R/a) = 1022208*(R/a)^6 - 735152*(R/a)^4\\
    &+ 178200*(R/a)^2 - 16875 = 0\\
\end{split}
\end{equation}

We find the real root of this polynomial equation by substituting $\zeta = (R/a)^2$, calculating our analytical upper bound of

\begin{equation}
\begin{split}
    R/a \le \frac{4177}{17424} - \frac{194471}{303595776 \sqrt[3]{\frac{6006250333}{5289852801024} + \frac{7675 \sqrt{3014}}{371061504}}}\\
    + \sqrt[3]{\frac{6006250333}{5289852801024} + \frac{7675 \sqrt{3014}}{371061504}} \approx 0.60522\\
\end{split}
\label{eq:f1Limit}
\end{equation}

\clearpage
\section{Pentagonal Stacking of Hard Tetrahedra}

\begin{figure}[h]
    \centering
    \includegraphics[width=0.65\linewidth]{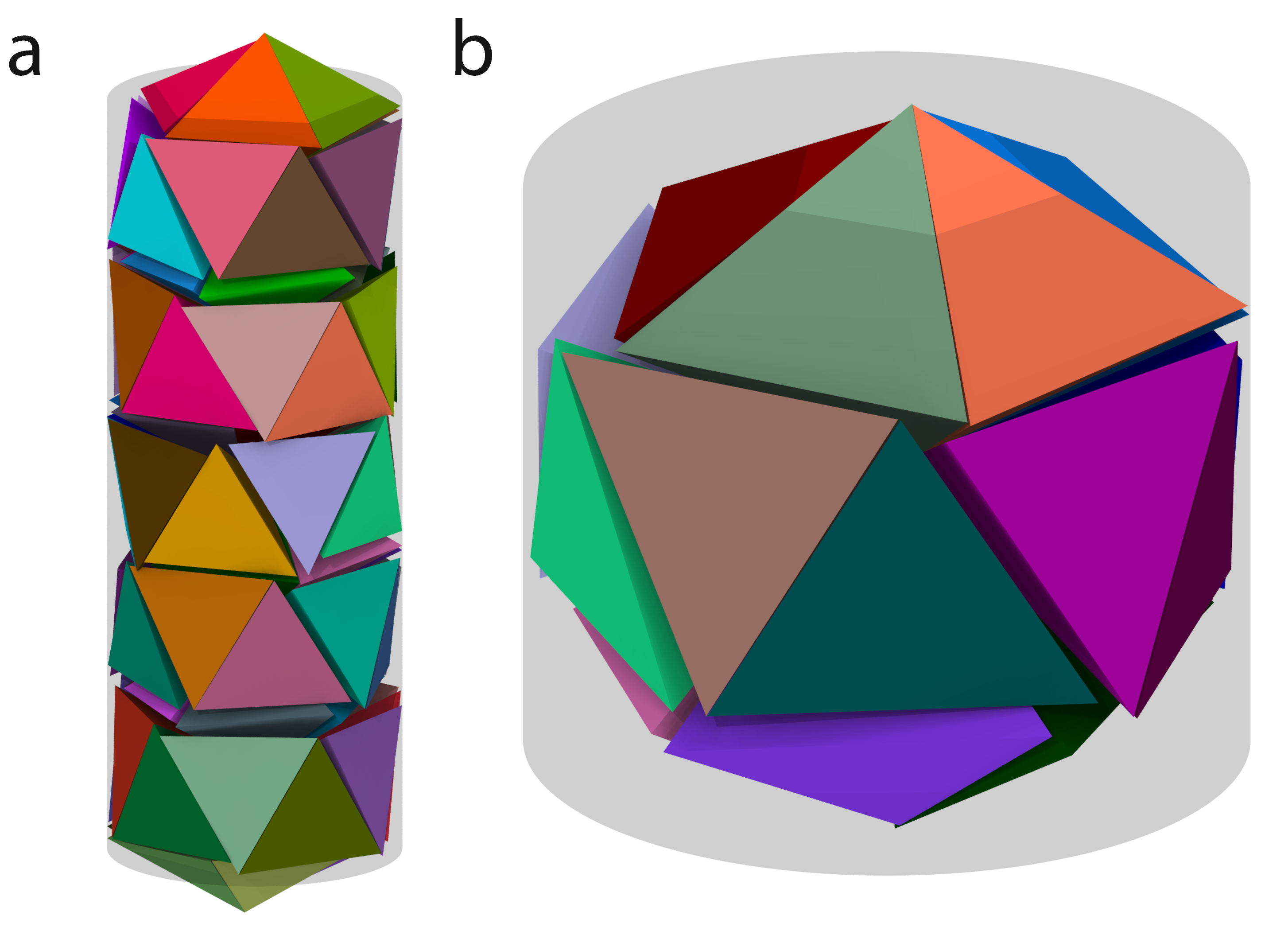}
    \caption{(a) Pentagonal columns, obtained from simulations with $N=15$, form the densest packing of hard tetrahedra in cylindrical confinement ($\phi=0.6409$) at $D/a=1.85529$. (b) Snapshot of the pentagonal core motif and the decagonal outer ring.}
    \label{fig:pentagonal_stacking}
\end{figure}

In simulations beyond the $D/a>1.21045$ limit, we observe pentagonal stackings as densest packings. In particular, we identify a dense structure with $\phi=0.6409$ at a diameter ratio of $D/a=1.85529$ containing $N=15$ particles. Here, the tetrahedra form a pentagonal core sandwiched between two decagonal outer rings, as illustrated in Fig.~\ref{fig:pentagonal_stacking}. Although hard tetrahedra in bulk also form pentagonal stacking motifs in the dodecagonal quasicrystal phase, the two structures differ in the number and arrangement of tetrahedra in the outer rings. The densest packing in cylindrical confinement features 10 particles per outer ring, whereas the corresponding motifs in the quasicrystal contain 12-fold rings. In addition, the rings in the quasicrystal columns are tilted with respect to one another, whereas the columns obtained here remain straight as a consequence of the cylindrical confinement.

\end{document}